\documentclass[twocolumn,showpacs,preprintnumbers,amsmath,amssymb,latexsym]{revtex4-1}
\usepackage{latexsym,amssymb,amsfonts,amsmath}
\usepackage{color}
\usepackage[dvips]{graphicx}

\newcommand{\msub}[1]{\ensuremath _{\mbox{\scriptsize #1}}}

\newcommand{\la}{\langle}

\newcommand{\ra}{\rangle}

\newcommand{\be}{\begin{equation}}
\newcommand{\ee}{\end{equation}}

\usepackage{graphicx}% Include figure files
\usepackage{dcolumn}% Align table columns on decimal point
\usepackage{bm}% bold math
\usepackage{ulem}

\begin{document}

%\preprint{APS/123-QED}

%\title{Universal properties of the QCD Dirac spectrum at high temperature}
\title{Anderson localization in QCD-like theories}
% Force line breaks with \\

\author{Matteo Giordano and Tam\'as G.\ Kov\'acs}
 %\altaffiliation[Also at ]{Physics Department, XYZ University.}%Lines break
 %automatically or can be forced with \\
\affiliation{%
Institute for Nuclear Research of the Hungarian Academy of Sciences \\
H-4026 Debrecen, Bem t\'er 18/c, Hungary %\textbackslash\textbackslash
}%

\author{Ferenc Pittler}
%\affiliation{Inst.\ for Theoretical Physics, E\"otv\"os University \\
%P\'azm\'any P.\ s\'et\'any 1/A, H-1117 Budapest, Hungary}

\affiliation{MTA-ELTE Lend\"ulet Lattice Gauge Theory Research Group \\
             P\'azm\'any P.\ s\'et\'any 1/A, H-1117 Budapest, Hungary}


\date{\today}% It is always \today, today,

\begin{abstract}

We review the present status of the Anderson transition in the spectrum of the
Dirac operator of QCD-like theories on the lattice. Localized modes at the
low-end of the spectrum have been found in SU(2) Yang-Mills theory with
overlap and staggered valence fermions as well as in $N_f=2+1$ QCD with
staggered quarks. We draw an analogy between the transition from localized to
delocalized modes in the Dirac spectrum and the Anderson transition in
electronic systems. The QCD transition turns out to be in the same
universality class as the transition in the corresponding Anderson model. We
also speculate on the possible physical relevance of this transition to QCD at
high temperature and the possible finite temperature phase transition in
QCD-like models with different fermion contents.

\end{abstract}

\maketitle

\section{\label{sec:Intro} Introduction}

Quantum Chromodynamics (QCD), the theory of strong interactions, is an
(apparently) extremely simple, yet very rich theory. Based on the
simple geometric principle of local gauge invariance, QCD has only the
gauge group and the quark masses as input parameters. Moreover, stable
matter around us is made of only $u$ and $d$ quarks that are almost
massless. In spite of this conceptual simplicity QCD produces a wealth
of nontrivial phenomena, such as generating most of the mass of
ordinary matter, quark confinement, the spontaneous breaking of chiral
symmetry and the anomalous breaking of the $U(1)_A$ symmetry. Recently
a new item has been added to this list, Anderson localization
\cite{Kovacs:2010wx}. 

Anderson localization was originally proposed to explain the loss of zero
temperature conductance as a result of impurities in a conducting solid
\cite{Anderson58}. It is essentially the spatial localization of electronic
wave functions due to quantum interference caused by the presence of impurities.
Since the original proposal, Anderson-type transitions have been demonstrated
with electromagnetic \cite{emwaves} and sound waves \cite{sound} as well as
ultracold atoms \cite{atoms}. All these phenomena occur on atomic scales, thus
it comes as a surprise that strong interactions, operating on vastly different
length and energy scales, are also capable of exhibiting a similar
phenomenon. 

Let us now briefly recall how Anderson localization appears in QCD. It is
well-known that at low temperature the lowest part of the spectrum of the QCD
Dirac operator is described by Random Matrix Theory (RMT) and the
corresponding quark states are delocalized over the whole four-dimensional
space-time volume \cite{Verbaarschot:2000dy}. This has been successfully
exploited to extract the low-energy constants of chiral perturbation theory
from lattice QCD simulations. In contrast, at high temperature, above the
chiral transition, the lowest part of the Dirac spectrum is drastically
different: it consists of localized eigenmodes, and the corresponding
eigenvalues are statistically independent and obey Poisson statistics. This
applies to modes up to a critical point, $\lambda_c$, that we call the
``mobility edge'', using the terminology of Anderson
transitions. Above that point the spectral statistics is again
described by RMT and the eigenmodes become extended. The mobility edge
has a strong temperature dependence, vanishing roughly at the
pseudo-critical temperature of the chiral and deconfining cross-over
and rapidly shifting upwards at higher temperatures
\cite{Kovacs:2012zq}. 

The localized-delocalized or Poisson-RMT transition in the QCD Dirac spectrum
appears to be similar to the Anderson transitions observed before. This
similarity turns out to go much deeper than a loose analogy. In fact, like
real Anderson transitions, the QCD transition also becomes singular in the
thermodynamic limit and its correlation length critical exponent is compatible
with that of the corresponding Anderson model \cite{Giordano:2013taa}. This
shows that the QCD transition is a real Anderson transition belonging to the
same universality class.  

The present paper is a summary of our current understanding of Anderson
localization in the high temperature, quark-gluon plasma phase of strongly
interacting matter as well as other QCD-like theories.  Besides an
introduction to the subject and a review of already published results, here we
also offer some new results and speculations as to the physical nature of the
transition.

\section{Lattice QCD and the Anderson model}

Since on hadronic energy scales QCD is a strongly interacting theory, it is
not amenable to perturbative methods. At low energies the only way to compute
physical quantities using systematically controllable approximations is to
discretize the theory on a four-dimensional Euclidean space-time lattice. The
discretized system has a finite number of degrees of freedom in a finite
physical volume and thus provides a regularization of the theory. In this
formulation the gauge fields are SU(3) valued parallel transporters attached
to the links of the hypercubic lattice, while quarks are represented by
Grassmann fields on the sites of the lattice.

The propagation of quarks is described by the covariant Dirac operator that
has several physically equivalent discretizations in common use. Here we will
mostly consider the simplest of these, the staggered Dirac operator
\cite{Susskind:1976jm}-\cite{Banks:1976ia}. Technically, it is a large sparse
anti-Hermitian matrix that has zeros in its diagonal and only the
matrix elements connecting nearest neighbor lattice sites (hopping
terms) are non-zero. These matrix elements depend on the gauge group
valued link variables. The gauge links themselves are random variables
with a distribution generated with the full path integral measure
\cite{Montvay:1994cy}. In this way the Dirac operator of lattice QCD
is a sparse random matrix, with purely imaginary eigenvalues
$i\lambda$.

For comparison, here it is useful to recall the most extensively studied model
of Anderson localization, the Anderson tight binding model. It describes
non-interacting electrons with the one-electron Hamiltonian
\be
 H =   \sum_i \varepsilon_i |i\rangle \langle i| \; + \;
       \sum_{(ij)} |i\rangle \langle j|, 
\ee
where the first summation is over the sites of the lattice while the second
one runs over all nearest neighbor pairs of sites. The states $|i\rangle$
represent atomic or molecular orbitals on the lattice sites with energies
$\varepsilon_i$ and the second sum contains the hopping terms. In the Anderson
model the $\varepsilon_i$ are i.i.d.\ random variables representing
disorder. The strength of the disorder is controlled by the width of the
distribution of $\varepsilon_i$-s.  In the special case when this is zero,
meaning that there is no disorder, the Anderson model is just the tight
binding model with delocalized Bloch-wave eigenmodes and the usual band
structure of the spectrum. If disorder is gradually turned on by increasing
the width of the distribution of the $\varepsilon_i$-s then localized states
appear at the band edges, while states at the band center still remain
delocalized. In the spectrum, localized and delocalized states are separated
by critical points called mobility edges. If the disorder strength increases,
the mobility edges move towards the band center and eventually at a critical
disorder the whole band becomes localized.

We have seen that both the Anderson Hamiltonian and the QCD Dirac operator can
be viewed as sparse random matrices with the non-zero Dirac matrix elements
representing hopping between nearest neighbor lattice sites.  Given the
similar structure of the Anderson Hamiltonian and the QCD Dirac operator, it
is natural to ask whether the QCD Dirac operator can also exhibit localization
phenomena similar to the one found in the Anderson model.  It has been known
for a long time that at low temperature the low-lying eigenmodes of the Dirac
operator are delocalized and the statistics of the corresponding part of the
spectrum is described by RMT. In the past two decades the random matrix theory
description of the low Dirac spectrum has been extensively studied and also
exploited to extract physical parameters from lattice simulations (see
\cite{Verbaarschot:2000dy} for an extensive review of the subject). 

Since the low-lying Dirac modes, the ones analogous to the band edge in the
Anderson model, are already delocalized, there seems to be little hope of
finding an Anderson transition in QCD. However, this applies only to the low
temperature hadronic phase of the system. In contrast, at high temperature,
above the chiral and deconfining transition, very little had been
known until recently about the nature of quark states at the lower
edge of the spectrum. This is all the more surprising since already
two decades ago it was suggested that the finite temperature chiral
transition might be accompanied by an Anderson-like transition in the
Dirac spectrum \cite{Halasz:1995vd}. This idea, however, had not been
followed up until more than a decade later when it was found in
lattice simulations that the chiral transition is indeed accompanied
by the appearance of more localized Dirac eigenmodes and a change 
in the spectral statistics towards Poisson type
\cite{GarciaGarcia:2006gr}. However, at that time a detailed  
verification of an Anderson type transition in QCD was still not
available.

\section{The Anderson transition in QCD}

The first quantitative demonstration of an Anderson-type transition in a
QCD-like theory came in a simplified model, SU(2) gauge theory with overlap
valence quarks \cite{GarciaGarcia:2006gr}. The overlap is a more complicated
discretization of the Dirac operator that, unlike other, simpler
discretizations, possesses an exact chiral symmetry
\cite{Narayanan:ss}-\cite{Neuberger:1997fp}. Therefore, it is particularly
suitable for any study focusing on the low-end of the Dirac spectrum. However,
it is rather expensive to simulate, therefore large volume simulations with
dynamical quarks, comparable to the ones that we will present here with
staggered quarks, are still impossible. In this simplified model, one of us
found that at high temperature the distribution of the lowest two Dirac
eigenvalues is precisely described by assuming Poisson eigenvalue statistics
\cite{Kovacs:2009zj}. This indirectly confirms in a quantitative manner that
at the given temperature the lowest part of the QCD Dirac spectrum is indeed
localized in the same way as in Anderson localization.

Unfortunately, overlap fermions provided to be too expensive to obtain enough
eigenvalues and statistics to reach up to the mobility edge and trace out the
transition between localized and delocalized modes within the spectrum in
detail. For the same reason we had to resort to the quenched approximation,
that is neglecting the quark determinant in the path integral measure. To
overcome these limitations, first two of us used staggered valence quarks to
trace out the transition from localized to delocalized modes through the
mobility edge within the spectrum. This was again a study at a fixed
temperature and using the quenched approximation with the SU(2) gauge group
\cite{Kovacs:2010wx}. 

Finally, two of us did the first detailed quantitative study of the Anderson
transition in full QCD without any major compromise
\cite{Kovacs:2012zq}. This again involved staggered quarks albeit including
$N_f=2+1$ flavors of sea quarks with their masses tuned to the physical
up/down and strange quark masses respectively.

\section{Level spacing statistics}

The simplest and most generally used method to locate a transition from
localized to delocalized states is by looking at statistical properties of the
corresponding spectrum. In the present section we study the simplest such
statistics, the unfolded level spacing distribution. This is a statistics that
can be defined {\it locally} anywhere in the spectrum, even for moderate sized
systems. Remarkably, this distribution is universal in the sense of being
independent of the physical details of the system, depending only on whether
states are localized or delocalized in the given spectral region.

States localized in distant spatial locations are statistically
independent. This is because fluctuations in the background disorder (gauge
field in QCD or the local potential in the Anderson model) influence only
those states that have a significant amplitude in the given
location. Localized states do not overlap and as a result, any fluctuation in
the background disorder can only change at most one of those states. Thinking
in terms of perturbation theory, even though nearby states are close in the
spectrum and the energy denominators are small, the matrix elements of local
operators between these states are small because of the large spatial
separation. Therefore, such states are not mixed by local
fluctuations. Statistical independence of such states means that their
distribution is Poissonian.

The other extreme possibility for the states is to be delocalized over the
whole system. In this case the spectral statistics is more complicated and 
it is described by Random Matrix Theory (RMT) (for a review, see
e.g.\ \cite{Guhr:1997ve}). For a quantitative analysis of the level spacing
distribution (LSD) it is necessary to perform a transformation on the spectrum,
called unfolding. Unfolding is essentially a local rescaling of the
eigenvalues to render the spectral density unity throughout the spectrum. The
unfolded level spacing distribution (ULSD) is the distribution of the quantity 
\be
  s = \frac{\lambda_{i+1} - \lambda_i}{\langle \lambda_{i+1} - \lambda_i
    \rangle},
\ee
where $\lambda_i$ are the ordered eigenvalues in a narrow spectral window and
the denominator is the average level spacing in the same spectral window. The
spectral window should be chosen narrow enough for the spectral statistics to
be constant in it but wide enough to include several eigenvalues in
the given volume. In principle this can always be fulfilled by
choosing a large enough volume since the spectral density is
proportional to the volume.
    
\begin{figure}
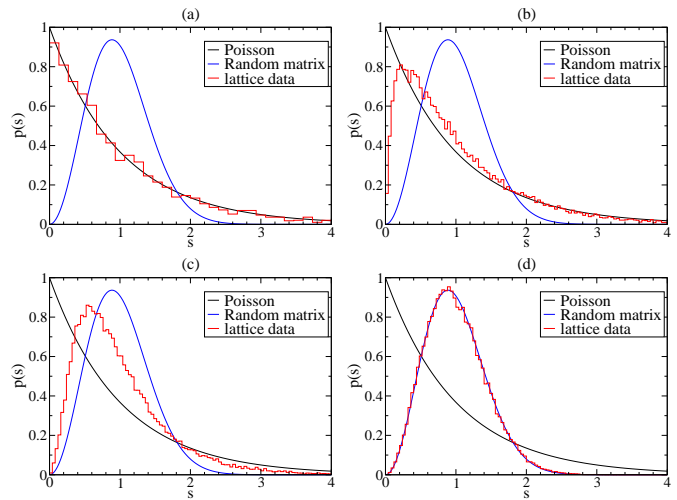

\begin{center}
\begin{tabular}{cc}
\includegraphics[width=0.50\columnwidth,keepaspectratio]{lsd_0.18-0.20.eps} & 
\includegraphics[width=0.50\columnwidth,keepaspectratio]{lsd_0.29-0.31.eps} \\
\includegraphics[width=0.50\columnwidth,keepaspectratio]{lsd_0.32-0.34.eps} &
\includegraphics[width=0.50\columnwidth,keepaspectratio]{lsd_0.37-0.386.eps} \\
\end{tabular}
\end{center}
\caption{\label{fig:ulsd}The unfolded level spacing distribution in
  different regions of the Dirac spectrum. The panels from (a) through (d)
  correspond to spectral windows going up in the spectrum.
%  the spectral regions $0.15 \leq \lambda a \leq 0.19$ (a), $0.29 \leq \lambda
%  a \leq 0.32$ (b), $0.34 \leq \lambda a \leq 0.35$ (c) and $0.375 \leq
%  \lambda a \leq 0.385$ (d). 
  The dashed line indicates the exponential distribution corresponding to the
  localized (Poisson) case and the dotted line indicates the chiral unitary
  Wigner surmise expected in the delocalized (RMT) case.}
\end{figure}

Unfolding eliminates non-universal features of the spectral statistics and the
resulting ULSD is fully universal. For localized eigenmodes and Poisson
statistics the ULSD is the standard exponential distribution,
\be 
  P(s) = \exp(-s).
   \label{eq:expd}
\ee
For delocalized modes the ULSD can still be computed analytically and it is
very precisely approximated by the Wigner surmise of the corresponding random
matrix universality class. QCD with staggered quarks in the fundamental
representation belongs to the unitary class and the corresponding Wigner
surmise is 
\be 
   P(s) = \dfrac{32}{\pi^{2}}s^{2}\cdot
                       \exp\left(-\dfrac{4}{\pi}s^{2}\right).
  \label{eq:wigners}
\ee
It is remarkable that these distributions do not have any free parameters and
thus a precise agreement of the ULSD with either of these is a strong
indication for the corresponding eigenmodes to be localized or delocalized
respectively. 

\begin{figure}
\begin{center}
\includegraphics[width=0.92\columnwidth,keepaspectratio]{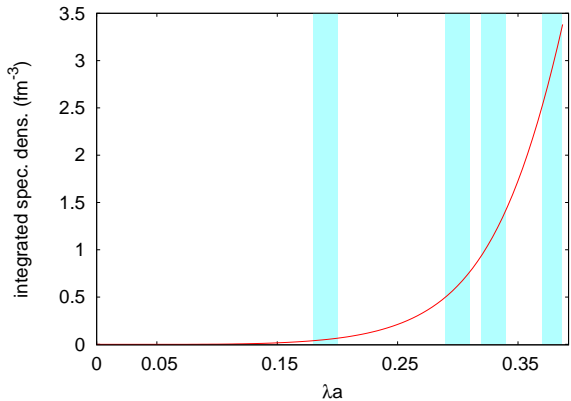}
\caption{\label{fig:spd48}  The integrated spectral density,
  normalized by the volume, in units of fm$^{-3}$ in the spectrum
  across the transition. The mobility edge is at $\lambda_c a=0.336$.
  The boxes correspond to the spectral windows of Fig.~\ref{fig:ulsd}.}
\end{center}
\end{figure}

To demonstrate the transition from localized to delocalized states within the
QCD Dirac spectrum, in Fig.~\ref{fig:ulsd} we show the ULSD in four different
spectral windows starting from the lowest part of the spectrum and going
upwards. The data is based on simulations with $N_f=2+1$ flavors of stout
smeared staggered quarks. The temporal lattice size was $N_t=4$ and
the lattice spacing $a=0.125$~fm, which corresponds to a temperature of
$T=400$~MeV. Details of the action and the scale setting can be found in
Ref.~\cite{Aoki:2005vt}. We also show the predictions for the ULSD coming from
Poisson statistics and from RMT in Eqs.\ (\ref{eq:expd}) and
(\ref{eq:wigners}) respectively. It is clear that starting from the lowest
modes and going upwards, the spectrum undergoes a transition from Poisson to
Wigner-Dyson statistics. It is interesting to compare this with how
the spectral density increases throughout the spectrum. In
Fig.~\ref{fig:spd48} we plot the integrated spectral density in the
spectrum through the transition.

\section{\label{sec:fss} 
Finite size scaling and the correlation length critical exponent}

The transition in the spectral statistics is a clear indication that a
localization-delocalization type transition takes place in the QCD Dirac
spectrum. On the other hand, there is a finite spectral window where the
statistics is clearly in between the two universal possibilities already
indicated. This is, however, not unexpected. Transitions in finite systems are
always smooth, and genuine critical behavior, exhibiting sharp transitions, can
only occur in the thermodynamic limit. One possibility to decide whether there
is such a transition for an infinite system is to study the dependence of the
transition on the system size using finite size scaling. In this section we
summarize the results of a finite size scaling study of the transition in the
QCD spectrum. For more technical details of this analysis we refer the reader
to Ref.~\cite{Giordano:2013taa}. 

To quantify the sharpness of the transition we choose a particular quantity
characterizing the unfolded level spacing distribution and monitor how it
changes through the transition in systems of various sizes. In principle, any
quantity that has different values for the limiting distributions of Eqs.\
(\ref{eq:expd}) and (\ref{eq:wigners}) would be suitable for this
purpose. A convenient choice is the integral of the probability density 
\be
  I = \int_0^{s_0} P(s) \, ds
   \label{eq:I}
\ee
to the point $s_0\approx 0.5$ where the two limiting distributions cross. For
our purposes $I$ is a good quantity since it decreases monotonically
from the exponential to the Wigner-Dyson distribution and the difference of
the values it takes in the limiting cases is large enough.  At the same time
$I$ receives contributions from a substantial fraction of the modes
following the exponential distribution.  This helps to improve the
quality of the observable at the low-end of the spectrum where the 
statistics is otherwise rather limited due to the small spectral
density there. 

\begin{figure}
\begin{center}
\includegraphics[width=0.9\columnwidth,keepaspectratio]{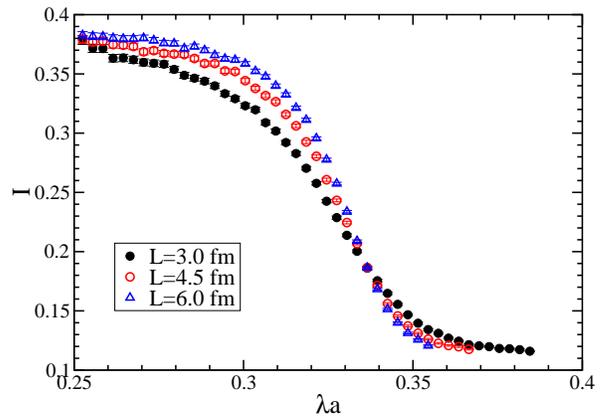}
\caption{\label{fig:Iu}  The quantity $I$ defined by Eq.\ (\ref{eq:I})
  computed in narrow spectral windows across the transition. The different
  symbols represent data taken from systems of different spatial sizes
  indicated in the legends.}
\end{center}
\end{figure}

In Fig.~\ref{fig:Iu} we show this quantity computed separately in different
spectral windows across the spectrum. The different symbols correspond to
linear spatial sizes of $L=3.0, 4.5$ and $6.0$~fm. The transition clearly
becomes sharper as the spatial volume of the system increases.

To obtain a more quantitative description of the dependence of the transition
on the linear size of the system, $L$, we use finite size scaling based on a
renormalization group (RG) argument. As is usual in Anderson transitions, we
assume a single relevant variable controlling the correlation length. In our
setup this can be taken to be $\lambda$, the location in the spectrum. To
simplify the notation, in the following discussion we write $\lambda$ instead
of $\lambda-\lambda_c$ where $\lambda_c$ is the mobility edge, that is the
critical point in the spectrum where the transition occurs. We have to keep in
mind that $\lambda_c$ is a parameter that also has to be determined from the
finite size scaling.

We assume that the quantity $I$ that we consider here depends on
$\lambda$, $L$ and some leading irrelevant variable that we call $\mu$. A
blocking transformation with scale $b>1$ transforms these parameters as
\be
  \lambda \rightarrow b^{1/\nu} \lambda, \hspace{2ex}
  L \rightarrow b^{-1} L, \hspace{2ex}
  \mu \rightarrow b^{y_\mu} \mu,
\ee
where $\nu>0$ is the correlation length critical exponent and $y_\mu<0$ is the
leading (smallest magnitude) irrelevant exponent and we used the linear
approximation of the blocking transformation around the given fixed
point. Other irrelevant operators are neglected. Since $I$ is a
dimensionless RG-invariant quantity it does not change under a scale $b$
blocking transformation. It means that
\be
   I(\lambda, \mu, L) \; = \;  I(b^{1/\nu}
   \lambda, b^{y_\mu} \mu, b^{-1} L). 
\ee
Choosing the blocking factor $b$ to be proportional to the system size,
$b=L/C$, systems of various sizes can be blocked down to the same
``reference size'' $C$ and compared. In this way the dependence of
$I$ on the system size through its last argument can be eliminated
and
\be
\begin{aligned}
  I(\lambda, \mu, L) \; &= \;
  I((CL)^{1/\nu} \lambda, (CL)^{y_\mu} \mu, C) 
  \; \\ &= \; f(L^{1/\nu} \lambda, L^{y_\mu} \mu),
\end{aligned}
\label{eq:irrelev}
\ee
where $f$ is a scaling function. Since $y_\mu$, being the exponent of an
irrelevant operator, is negative, in principle the size of the system can be
chosen large enough for $L^{y_\mu} \mu \approx 0$ and then
$I$ depends only on the single variable $L^{1/\nu} 
\lambda$. Note that since the correlation length in an infinite system is
proportional to $\lambda^{-\nu}$, this single independent variable
is essentially the ratio of the system size and the correlation length.

Singular behavior is not expected in a finite system and as a result the
dependence of $I$ on $\lambda$ is analytic and can be expanded as 
\be
 I(\lambda, L) = F(L^{1/\nu} (\lambda-\lambda_c)) = 
   \sum_n F_n L^\frac{n}{\nu} (\lambda - \lambda_c)^n,
    \label{eq:fss}
\ee
where we have restored the explicit dependence on the critical point,
$\lambda_c$.

\begin{figure}
\begin{center}
\vspace{4ex}
\includegraphics[width=0.9\columnwidth,keepaspectratio]{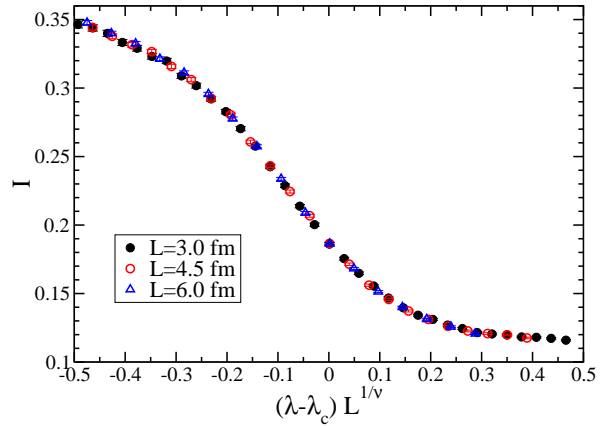}
\caption{\label{fig:Iu_scal} The quantity $I$ defined by
  Eq.\ (\ref{eq:I}) computed in narrow spectral windows across the
  transition. The different symbols represent data taken from systems of
  different spatial sizes indicated in the legends. The data is plotted
  against the scaling variable $L^{1/\nu} (\lambda - \lambda_c)$ with the
  parameters $\nu$ and $\lambda_c$ obtained from the fit (see text).}
\end{center}
\end{figure}

Finite size scaling relies on the first equality of Eq.\ (\ref{eq:fss}).  This
means that data for $I$ computed on systems of different sizes, if
plotted as a function of the appropriate scaling variable $L^{1/\nu}
(\lambda-\lambda_c)$, should all collapse on a single scaling curve . Our task
is to determine the parameters $\lambda_c$ and $\nu$ resulting in such data
collapse (see Fig.\ \ref{fig:Iu_scal}). The simplest way of doing that
is by making use of the expansion in Eq.\ (\ref{eq:fss}), truncated to
an appropriate order, $n$. Data taken on different volumes can be
fitted to this form to extract the parameters $\lambda_c, \nu$ and
$F_0, F_1,\ldots,F_n$. The truncation order, $n$, should be large enough
to properly describe the scaling function, $F$, in the fitting range
but small enough to ensure stability of the fits.

\begin{figure}
\begin{center}
\includegraphics[width=0.9\columnwidth,keepaspectratio]{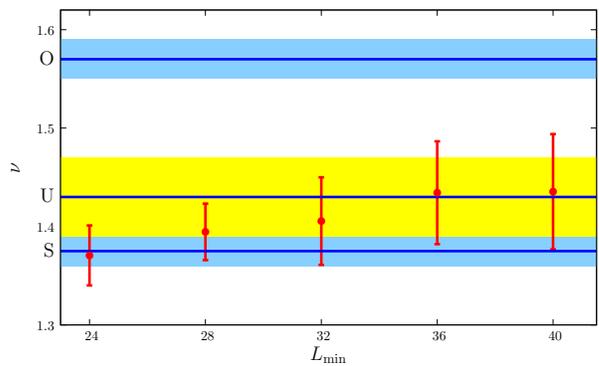}
\caption{\label{fig:nu_vs_Lmin} (Reprinted from Ref.\ \cite{Giordano:2013taa})
  The fitted value of the critical exponent $\nu$ versus $L\msub{min}$, the
  linear size of the smallest system for which data was included in the
  fit. The three horizontal lines and shaded regions represent the critical
  exponents and their uncertainties for the Anderson models of orthogonal,
  unitary and symplectic symmetry classes (marked from top to bottom and
  obtained from Refs.~\cite{corrections,nu_unitary,nu_symp}).}
\end{center}
\end{figure}

This procedure gives correct results only if the volumes included in the fit
are large enough that corrections coming from the neglected irrelevant
parameter(s) (see Eq.\ (\ref{eq:irrelev})) are small. Such corrections can be
systematically accounted for by replacing Eq.\ (\ref{eq:fss}) with a double
expansion containing also the dependence of $I$ on the leading
irrelevant variable, $\mu$. However, this makes the number of parameters to be
fitted so large that very good quality data is needed to ensure a stable
fit. For this reason we followed a different procedure. We did the fit using
only subsets of the full data by omitting data for system sizes smaller than
$L\msub{min}$. In Fig.~\ref{fig:nu_vs_Lmin} we show how the fitted value of
the exponent $\nu$ depends on the smallest system size that is included in the
fit. Naturally, if less data is used for the fit, its uncertainty increases
but the fitted value stabilizes, showing that corrections to the assumed
one-parameter scaling of Eq.~(\ref{eq:fss}) are not larger than the other
sources of uncertainty. It is remarkable that the critical exponent we obtain
for the QCD transition is consistent with that of the unitary Anderson
model. This strongly indicates the these two seemingly very different models
are in the same universality class.

\section{Continuum limit and the temperature dependence of the mobility edge}

We have demonstrated that the transition in the QCD Dirac spectrum is a
genuine Anderson transition belonging to the same universality class as the
Anderson model of the corresponding (unitary) symmetry class. However, this
was done only at a fixed lattice spacing. Since the physical theory is defined
only as the continuum limit of lattice QCD, it is important to check how the
transition changes in the continuum limit. In this respect the most important
quantity to look at is the mobility edge, $\lambda_c$. In finite temperature
lattice simulations the temperature is controlled by the system size in the
temporal direction as
\be
  T = \frac{1}{N_t a},
\ee
where $N_t$ is the temporal size of the box in lattice units and $a$ is the
lattice spacing. Since $N_t$ is typically a small integer, if the lattice
spacing is fixed, the temperature can be changed only in discrete
steps. For this reason, in Ref.~\cite{Kovacs:2012zq} we studied both
the temperature and the lattice spacing dependence of the mobility
edge using the same set of simulations. For this we combined three
different lattice spacings, $a=0.125, 0.082$ and $0.062$~fm with
$N_t=4,6,8$ (not all combinations were used). The parameters were
chosen so that at $T=400$~MeV we had simulations with all three
lattice spacings and the temperatures spanned the range between
260~MeV and 800~MeV. 

\begin{figure}
\begin{center}
%\vspace{4ex}
\includegraphics[width=0.9\columnwidth,keepaspectratio]{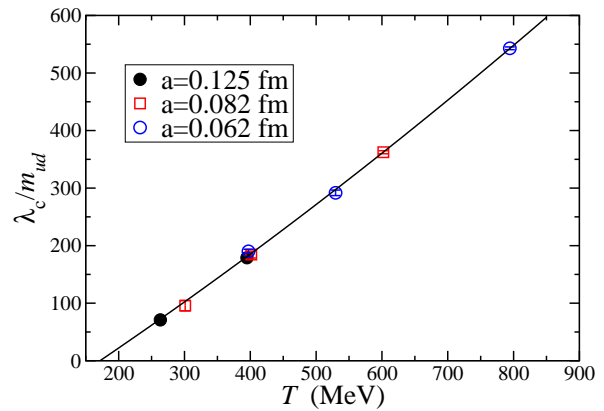}
\caption{\label{fig:lcpmvsT} (Reprinted from Ref.\ \cite{Kovacs:2012zq}) The
  mobility edge normalized by the light quark mass versus the
  temperature. Different symbols represent data originating from simulations
  with different lattice spacings. The continuous line is a quadratic fit to
  all the data.}
\end{center}
\end{figure}

The mobility edge is a dimensionful quantity and it is not a renormalization
group invariant. Since $\lambda_c$ characterizes the Dirac spectrum, it is
expected to be renormalized as the quark mass (see
Refs.~\cite{Giusti:2008vb} and \cite{Kovacs:2012zq} for a
discussion). For this reason the quantity that we considered was 
$\lambda_c/m\msub{ud}$, the mobility edge normalized by the light quark
mass. In Fig.~\ref{fig:lcpmvsT} we show the temperature dependence of this
quantity. Since the data obtained at different lattice spacings are all on a
smooth curve, we can conclude that scaling violations are small and the
Anderson transition also takes place in the continuum limit. 

Also in the same figure we show a quadratic fit of the form 
\be
 \frac{\lambda_c}{m\msub{ud}}(T) = a \cdot \frac{T-T_c}{T_c} + 
   b \cdot \left( \frac{T-T_c}{T_c} \right)^2
    \label{eq:Tc}
\ee
By construction, $T_c$ is the temperature where the mobility edge vanishes
and, as a result, the Anderson transition ceases to exist. The fit yields
$T_c=171(9)$~MeV which is consistent with the known location of the cross-over
\cite{Borsanyi:2010bp,Bazavov:2011nk}. We also note that in Eq.\ (\ref{eq:Tc})
the dimensionless coefficient of the quadratic term coming from the fit is two
orders of magnitude smaller than that of the linear term. The mobility
edge thus increases sharply with the temperature and, to a good
approximation, it scales linearly with the temperature.

\section{Shape analysis}

\begin{figure}[t]
  \centering
  \includegraphics[width=0.9\columnwidth]{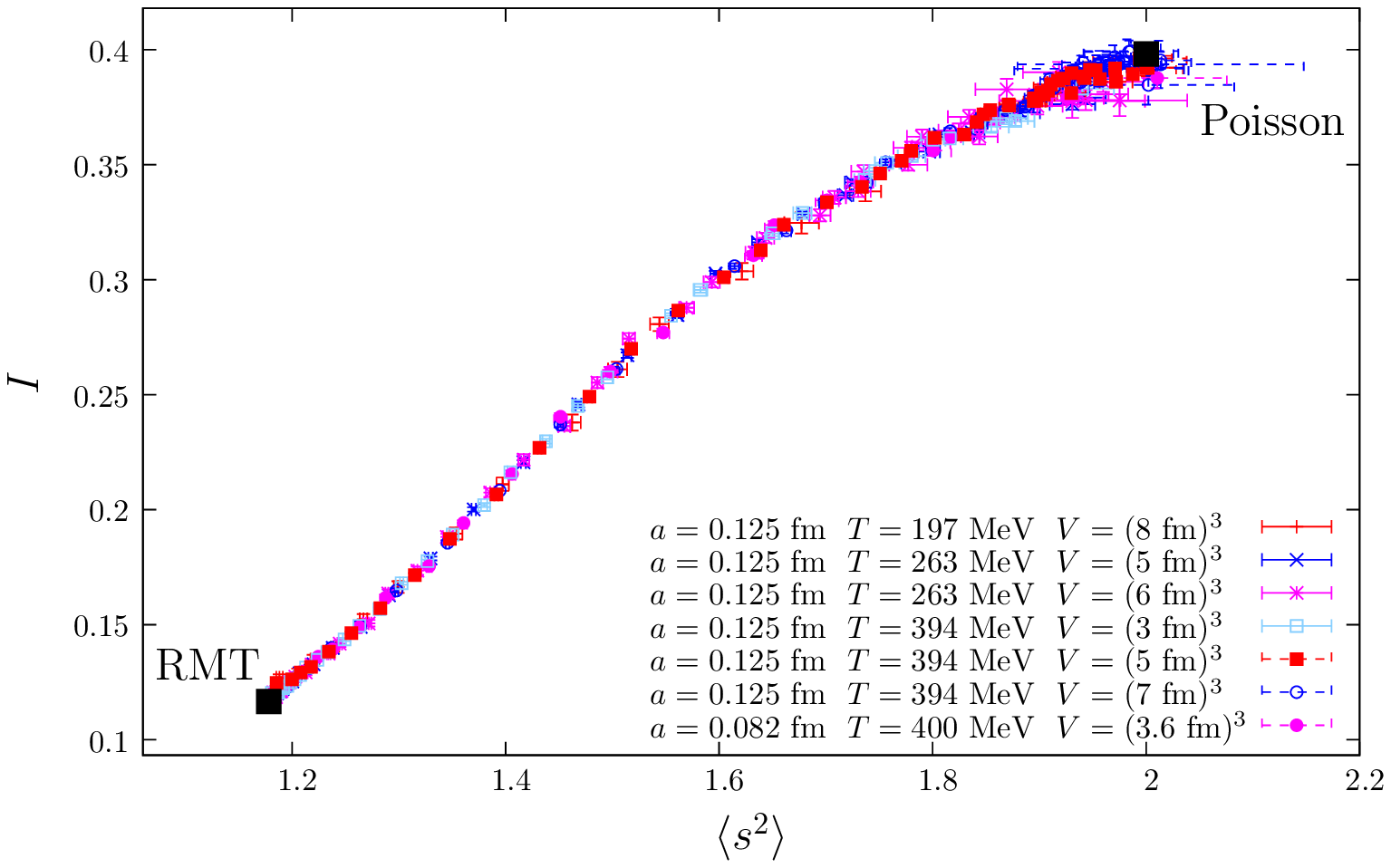}

  \vspace{2ex}
  \includegraphics[width=0.9\columnwidth]{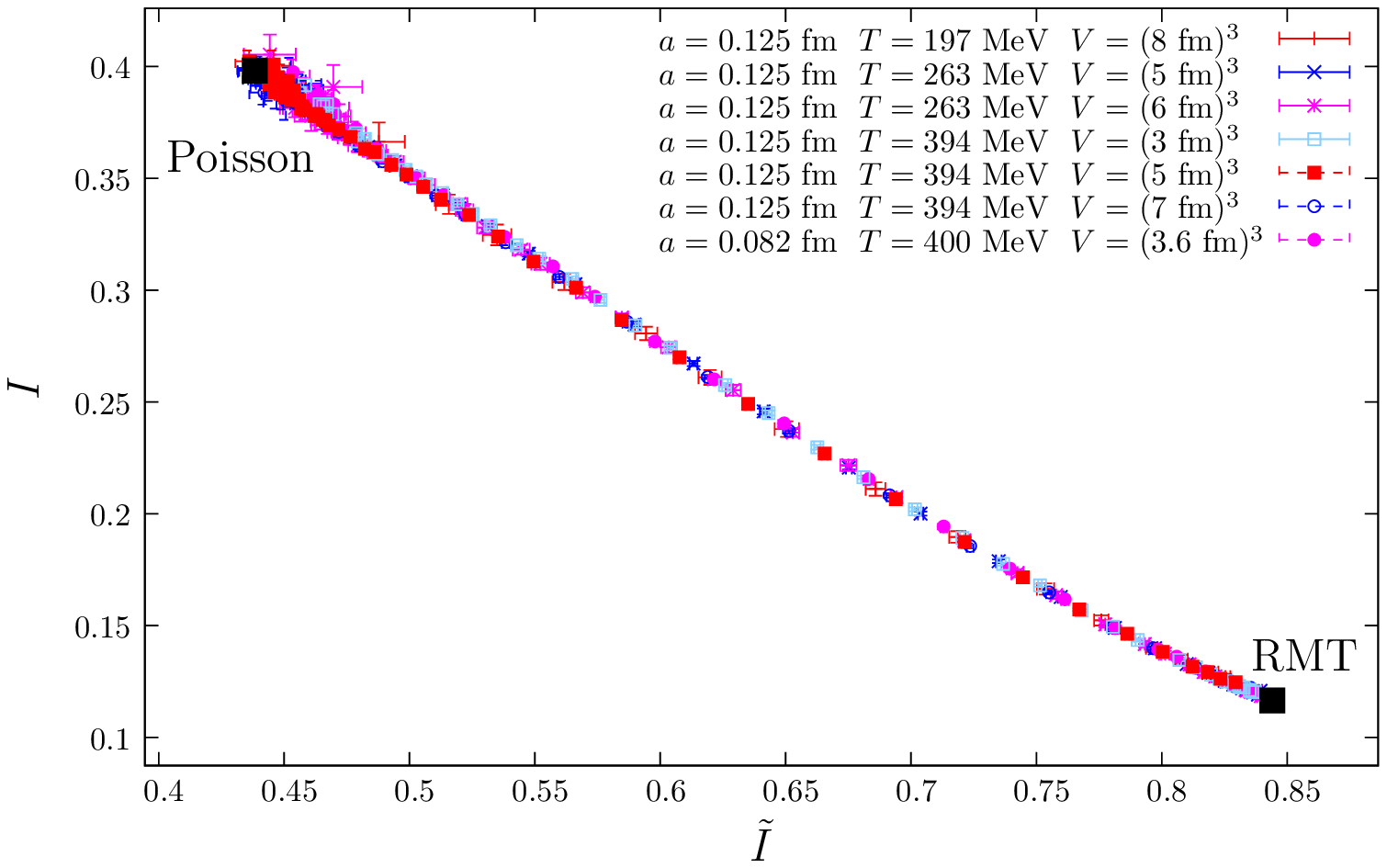}
  \caption{Shape analysis using $I$ (Eq. (\ref{eq:I}))
    and the second moment of the ULSD $\la s^2 \ra$ (top), and $I$ 
    and $\tilde I$ (Eq. (\ref{eq:tI})) (bottom).} 
  \label{fig:shape}
\end{figure}

In Section \ref{sec:fss} we analyzed in detail how $I$ changed
in the spectrum through the Anderson transition. Our finite size scaling
analysis was essentially based on a matching of points in the spectra of
systems of different sizes where $I$ took the same value. That such a
matching was possible by a simple transformation $\lambda \rightarrow
L^{1/\nu} (\lambda-\lambda_c)$ was indirect evidence that $I$
depended only on the ratio of the system size to the correlation length. This
is natural to expect in the case of an Anderson transition since there is only
one relevant variable around the fixed point characterizing the transition.  

The quantity $I$ is just one parameter characterizing the
ULSD. Through the transition the ULSD changes from the exponential
distribution to the Wigner surmise in a 
continuous way. Thus the ULSD traverses a path in the infinite dimensional
space of all possible distributions. An interesting question is how this path
depends on the particular details of the system such as the volume, the
lattice spacing and the physical temperature. A simple way to study this,
known as shape analysis \cite{shape_analysis}, is to compute different
parameters of the ULSD and plot them one against another. If the ULSD of
different systems follows the same universal path, the data coming from these
systems should collapse on a single universal curve. This curve can be viewed
as a two dimensional projection of the path that the ULSD traces. 

In Fig.~\ref{fig:shape} we show the results of the shape analysis for
two pairs of observables, namely $I$ and the second moment
of the ULSD, and $I$ and the integrated ULSD $\tilde I$,
defined as 
\be
\tilde I= \int_{s_0}^{s_1} ds\, P_\lambda(s)\,,
\label{eq:tI}
\ee
where $s_1\approx 1.8$ is the second crossing point of the exponential
and the Wigner surmise. Data obtained with different lattice size,
lattice spacing and/or temperature all fall on a universal curve, with
deviations being smaller than the statistical errors for both pairs of
observables. From this analysis one can conclude that the transition
from Poisson to Wigner-Dyson statistics takes place on a universal
path, up to small corrections. In turn, this hints to the local
spectral statistics depending on the position in the spectrum and on
the details of the system through a single quantity $\zeta$, i.e., the 
local ULSD $P=P(s;\lambda;L,a,T)$ is a function of the form
$P(s;\lambda;L,a,T)=p(s;\zeta(\lambda,L,a,T))$, up to small
corrections vanishing in the limit $L\to\infty$. In other words, the
local ULSD is essentially determined 
by a single physical quantity, which is most likely a property of the
corresponding eigenvectors, regardless of the details of the system. A
natural candidate is the correlation length  
%(properly defined) localisation length 
of the eigenvectors, which certainly is the appropriate quantity in
the vicinity of the critical point, where one-parameter scaling
applies. 

The universal path in the space of probability distributions
corresponds to a universal one-parameter family of random matrix
models, describing the transition from Poisson to Wigner-Dyson
behavior in the Dirac spectrum independently of the details of the
system. A comparison with analogous results obtained in the Anderson
model shows that the spectral statistics at the critical point in the
two models are compatible~\cite{Nishigaki}. This provides further
support to the claim that the localization/delocalization transitions
found in the two models belong to the same universality class.

\section{Conclusions}

We studied the transition from localized to delocalized states in the QCD
Dirac spectrum at high temperature. We demonstrated that the transition is a
genuine Anderson transition and that its critical exponent agrees with that of
the corresponding Anderson model. However, there is an important difference
between Anderson transitions in electronic systems and the transition in
QCD. On the one hand, in electronic systems the Anderson transition is a
genuine phase transition with the zero temperature conductivity having a
singularity. On the other hand, in QCD, most likely there is no singular
thermodynamic behavior associated with the Anderson transition. In fact, the
finite temperature transition from the hadronic to the quark-gluon plasma
state is known to be a cross-over \cite{Aoki:2006we}. The resolution of this
apparent paradox is that the QCD Dirac spectrum has no such direct physical
interpretation as the spectrum of the one-electron Hamiltonian. Neither the
appearance of low-lying localized states around $T_c$ nor the presence of a
mobility edge above $T_c$ manifests itself in singular thermodynamic
behavior. This is because there is no thermodynamic quantity in QCD that is
sensitive enough to the abrupt change occurring at $\lambda_c$ in the
spectrum.

However, this might not be the case for some other QCD-like theories. It is
believed that if the quark masses were small enough the finite temperature
cross-over would become a phase transition \cite{pisarski-wilczek}. In that
case the extension of the singular line $\lambda_c(T)$ in the ``phase
diagram'' of Fig.~\ref{fig:lcpmvsT} to $T_c$ might contain useful information
about the chiral phase transition. 

Even though in QCD the Anderson transition does not correspond to a genuine
phase transition in observable quantities, the appearance of localized
states in the Dirac spectrum can have important physical
consequences. Correlators of hadronic operators can be written in
terms of quark propagators which in turn admit a decomposition into
Dirac eigenmodes. The eigenmodes in this decomposition are weighted with 
$(m+i\lambda)^{-1}$ where $m$ is the quark mass and $\lambda$ is the
eigenvalue. Low-lying Dirac modes, therefore, give a large contribution to
correlators. However, they can give a significant contribution to the
correlators only on distance scales smaller than their localization
length. Therefore, long-distance correlators are dominated by Dirac eigenmodes
above the mobility edge. In this way the mobility edge plays the role of an
effective gap in the spectrum, similar to the quark mass. In this connection
it is remarkable that this gap increases sharply with the temperature and even
around $T=2T_c$ it is already two orders of magnitude larger than the light
quark mass (see Fig.~\ref{fig:lcpmvsT}). Asymptotic hadronic correlators at
and above this temperature behave qualitatively as if the effective quark mass
were comparable to $\lambda_c$. This might provide an explanation of the sharp
increase of screening masses that was observed in lattice calculations above
$T_c$ \cite{Cheng:2010fe}.

There are several directions in which further study can provide interesting
new insight. Firstly, the full physical implications of localized Dirac
modes have certainly not been understood. The possible connections, if any, of
the QCD Anderson transition to finite temperature phase transitions in
QCD-like theories still remains to be explored. Secondly, there is much more
to be learned from a detailed study of how the quark eigenmodes themselves
change through the transition. The only information we have about this so far is
through the inverse participation ratio (IPR). Using the IPR we estimated that
the localization length of the localized modes is controlled by the inverse
temperature \cite{Kovacs:2012zq}. In Anderson transitions, however, it is
known that the critical wave functions develop a peculiar multifractal
structure that has recently been utilized to improve the finite size scaling
analysis \cite{multifractal}. It would be certainly interesting to explore
this possibility in the QCD Anderson transition.

Both the Anderson model and the lattice QCD Dirac operator are sparse random
matrices with a structure reflecting the geometry of space (or space-time). In
this context it would be interesting to understand what are the necessary
conditions of an Anderson-type transition in the spectrum of such a
system. One might be tempted to speculate that it is the spectral density per
unit volume that drives the transition.  Indeed, a small spectral density per
unit volume is almost surely a necessary condition for localized modes to
appear, as otherwise fluctuations of the gauge field are likely to mix the
modes due to their small energy differences.  However, that this is certainly
not the full story is shown by the example of QCD-like theories with many
fermion flavors. Introducing enough fermion flavors can drive those systems
into the deconfined chirally restored phase already at zero temperature; as a
result, the spectral density at the low-end of the spectrum becomes
arbitrarily small. However, no Anderson transition takes place in those systems
\cite{Landa-Marban:2013oia,Bietenholz:2013coa}. This example also shows that
there is much to be understood about Anderson transitions in QCD-like
theories.

Finally, we would like to remark that phenomena apparently similar to the one
described here were previously seen in {\it quenched} QCD at {\it zero
  temperature}. Golterman and Shamir found that just outside the Aoki phase
there is a finite density of localized Dirac modes related to so called
exceptional configurations, topological charge fluctuations on the scale of
the lattice spacing \cite{Golterman:2003qe}. However, these objects are very
different from the ones that we studied here, as they are not driven by the
temperature and are expected to be absent in the full theory with light
dynamical quarks. Greensite {\it et al.\/} observed that in zero
temperature quenched gauge backgrounds the covariant Laplacian also
possesses localized eigenmodes at the edge of its spectrum
\cite{Greensite:2005yu}. Also in this case there is no indication that
this phenomenon could be related to the transition in the spectrum at
finite temperature described in the present paper.

\section*{Acknowledgments}

TGK and MG are supported by the Hungarian Academy of Sciences under
``Lend\"ulet'' grant No.\ LP2011-011. We thank the Budapest-Wuppertal group
for allowing us to use their computer code for generating the lattice
configurations. Finally we also thank S. D.\ Katz and D.\ N\'ogr\'adi for
discussions.

%%%%%


\begin{thebibliography}{99}

%\cite{Kovacs:2010wx}
\bibitem{Kovacs:2010wx} 
  T.~G.~Kov\'acs and F.~Pittler,
  %``Anderson Localization in Quark-Gluon Plasma,''
  Phys.\ Rev.\ Lett.\  {\bf 105}, 192001 (2010)
  [arXiv:1006.1205 [hep-lat]].
  %%CITATION = ARXIV:1006.1205;%%

%\cite{Anderson58}
\bibitem{Anderson58}
P.~W.~Anderson, Phys.\ Rev.\ {\bf 109}, 1492 (1958).

%\cite{emwaves}
\bibitem{emwaves} 
  R.~Dalichaouch, J.P.~Armstrong, S.~Schultz, P.M.~Platzman, and S. L.~McCall, 
  %``Microwave localization by two-dimensional random scattering. ''
  Nature {\bf 354}, 53 (1991).

%\cite{sound}
\bibitem{sound}
  H.~Hu, A.~Strybulevych, J.H.~Page, S.E.~Skipetrov, and B.A.~Van Tiggelen,
  %``Localization of ultrasound in a three-dimensional elastic network'' 
  Nature Physics {\bf 4}, 945 (2008).

%\cite{atoms}
\bibitem{atoms}
  J.~Billy {\it et al.},
  %`` Direct observation of Anderson localization of matter waves in a
  %controlled disorder.'' 
  Nature {\bf 453}, 891 (2008);
%
  G.~Roati {\it et al.},
  %``Anderson localization of a non-interacting bose-einstein condensate''
  Nature {\bf 453}, 895 (2008).

%\cite{Verbaarschot:2000dy}
\bibitem{Verbaarschot:2000dy}
  J.~J.~M.~Verbaarschot and T.~Wettig,
  %``Random matrix theory and chiral symmetry in QCD,''
  Ann.\ Rev.\ Nucl.\ Part.\ Sci.\  {\bf 50}, 343 (2000)
  [arXiv:hep-ph/0003017].
  %%CITATION = ARNUA,50,343;%%

%\cite{Kovacs:2012zq}
\bibitem{Kovacs:2012zq} 
  T.~G.~Kov\'acs and F.~Pittler,
  %``Poisson to Random Matrix Transition in the QCD Dirac Spectrum,''
  Phys.\ Rev.\ D {\bf 86}, 114515 (2012)
  [arXiv:1208.3475 [hep-lat]].
  %%CITATION = ARXIV:1208.3475;%%

%\cite{Giordano:2013taa}
\bibitem{Giordano:2013taa} 
  M.~Giordano, T.~G.~Kov\'acs and F.~Pittler,
  %``Universality and the QCD Anderson Transition,''
  Phys.\ Rev.\ Lett.\  {\bf 112}, 102002 (2014)
  [arXiv:1312.1179 [hep-lat]].
  %%CITATION = ARXIV:1312.1179;%%

%\cite{Susskind:1976jm}
\bibitem{Susskind:1976jm} 
  L.~Susskind,
  %``Lattice Fermions,''
  Phys.\ Rev.\ D {\bf 16}, 3031 (1977).
  %%CITATION = PHRVA,D16,3031;%%

%\cite{Kogut:1974ag}
\bibitem{Kogut:1974ag} 
  J.~B.~Kogut and L.~Susskind,
  %``Hamiltonian Formulation of Wilson's Lattice Gauge Theories,''
  Phys.\ Rev.\ D {\bf 11}, 395 (1975).
  %%CITATION = PHRVA,D11,395;%%

%\cite{Banks:1976ia}
\bibitem{Banks:1976ia} 
  T.~Banks {\it et al.}  [Cornell-Oxford-Tel Aviv-Yeshiva Collaboration],
  %``Strong Coupling Calculations of the Hadron Spectrum of Quantum
  %Chromodynamics,''
  Phys.\ Rev.\ D {\bf 15}, 1111 (1977).
  %%CITATION = PHRVA,D15,1111;%%





%\cite{Montvay:1994cy}
\bibitem{Montvay:1994cy}
  I.~Montvay and G.~M\"unster,
  ``Quantum fields on a lattice,'' Cambridge, UK: Univ. Pr. (1994) 491
  p. (Cambridge monographs on mathematical physics).

%\cite{Halasz:1995vd}
\bibitem{Halasz:1995vd} 
  A.~M.~Hal\'asz and J.~J.~M.~Verbaarschot,
  %``Universal fluctuations in spectra of the lattice Dirac operator,''
  Phys.\ Rev.\ Lett.\  {\bf 74}, 3920 (1995)
  [hep-lat/9501025].
  %%CITATION = HEP-LAT/9501025;%%

%\cite{GarciaGarcia:2006gr}
\bibitem{GarciaGarcia:2006gr}
  A.~M.~Garc\'ia-Garc\'ia and J.~C.~Osborn,
  %``Chiral phase transition in lattice QCD as a metal-insulator transition,''
  Phys.\ Rev.\  D {\bf 75}, 034503 (2007)
  [arXiv:hep-lat/0611019].
  %%CITATION = PHRVA,D75,034503;%%


%\bibitem{overlap}

\bibitem{Narayanan:ss}
R.~Narayanan and H.~Neuberger,
%``Chiral Fermions On The Lattice,''
Phys.\ Rev.\ Lett.\  {\bf 71} (1993) 3251
[arXiv:hep-lat/9308011].
%%CITATION = HEP-LAT 9308011;%%

\bibitem{Narayanan:1994sk}
R.~Narayanan and H.~Neuberger,
%``Chiral determinant as an overlap of two vacua,''
Nucl.\ Phys.\ B {\bf 412}, 574 (1994)
[arXiv:hep-lat/9307006].
%%CITATION = HEP-LAT 9307006;%%

\bibitem{Narayanan:1995gw}
R.~Narayanan and H.~Neuberger,
%``A Construction of lattice chiral gauge theories,''
Nucl.\ Phys.\ B {\bf 443}, 305 (1995)
[arXiv:hep-th/9411108].
%%CITATION = HEP-TH 9411108;%%

\bibitem{Neuberger:1997fp} 
  H.~Neuberger,
  %``Exactly massless quarks on the lattice,''
  Phys.\ Lett.\ B {\bf 417}, 141 (1998)
  [hep-lat/9707022].
  %%CITATION = HEP-LAT/9707022;%%

%\cite{Kovacs:2009zj}
\bibitem{Kovacs:2009zj}
  T.~G.~Kov\'acs,
  %``Absence of correlations in the QCD Dirac spectrum at high temperature,''
  Phys.\ Rev.\ Lett.\  {\bf 104}, 031601 (2010)
  [arXiv:0906.5373 [hep-lat]].
  %%CITATION = ARXIV:0906.5373;%%

%\cite{Guhr:1997ve}
\bibitem{Guhr:1997ve} 
  T.~Guhr, A.~M\"uller-Groeling and H.~A.~Weidenmuller,
  %``Random matrix theories in quantum physics: Common concepts,''
  Phys.\ Rept.\  {\bf 299}, 189 (1998)
  [cond-mat/9707301].
  %%CITATION = COND-MAT/9707301;%%

%\cite{Aoki:2005vt}
\bibitem{Aoki:2005vt} 
  Y.~Aoki, Z.~Fodor, S.~D.~Katz and K.~K.~Szab\'o,
  %``The Equation of state in lattice QCD: With physical quark masses towards
  %the continuum limit,''
  JHEP {\bf 0601}, 089 (2006)
  [hep-lat/0510084];
  %%CITATION = HEP-LAT/0510084;%%

\bibitem{corrections}
K.~Slevin and T.~Ohtsuki, Phys.\ Rev.\ Lett.\ {\bf 82}, 382 (1999).

\bibitem{nu_unitary}
K.~Slevin and T.~Ohtsuki, Phys.\ Rev.\ Lett.\ {\bf 78}, 4083 (1997).

\bibitem{nu_symp} Y.~Asada, K.~Slevin and T.~Ohtsuki,
  {J. Phys. Soc. Jpn.} {\bf 74} supplement, 238 (2005)
  [cond-mat/0410190 [cond-mat.dis-nn]].

%\cite{Giusti:2008vb}
\bibitem{Giusti:2008vb} 
  L.~Giusti and M.~L\"uscher,
  %``Chiral symmetry breaking and the Banks-Casher relation in lattice QCD
  %with Wilson quarks,'' 
  JHEP {\bf 0903}, 013 (2009)
  [arXiv:0812.3638 [hep-lat]].
  %%CITATION = ARXIV:0812.3638;%%

%\cite{Borsanyi:2010bp}  Tc
\bibitem{Borsanyi:2010bp} 
  S.~Bors\'anyi {\it et al.}  [Wuppertal-Budapest Collaboration],
  %``Is there still any T_c mystery in lattice QCD? Results with physical
  %masses in the continuum limit III,'' 
  JHEP {\bf 1009}, 073 (2010)
  [arXiv:1005.3508 [hep-lat]];
  %%CITATION = ARXIV:1005.3508;%%
     
%\cite{Bazavov:2011nk} Tc
\bibitem{Bazavov:2011nk} 
  A.~Bazavov, T.~Bhattacharya, M.~Cheng, C.~DeTar, H.~T.~Ding, S.~Gottlieb,
  R.~Gupta and P.~Hegde {\it et al.}, 
  %``The chiral and deconfinement aspects of the QCD transition,''
  Phys.\ Rev.\ D {\bf 85}, 054503 (2012)
  [arXiv:1111.1710 [hep-lat]].
  %%CITATION = ARXIV:1111.1710;%%

\bibitem{shape_analysis} 
I.~Varga, E.~Hofstetter, M.~Schreiber and J.~Pipek, Phys.\ Rev.\ B
{\bf 52}, 7783 (1995). 

\bibitem{Nishigaki} S.M.~Nishigaki, M.~Giordano, T.G.~Kov\'acs and
    F.~Pittler, PoS  LATTICE {\bf 2013}, 018 (2013).

%\cite{Aoki:2006we}
\bibitem{Aoki:2006we} 
  Y.~Aoki, G.~Endr\H odi, Z.~Fodor, S.~D.~Katz and K.~K.~Szab\'o,
  %``The Order of the quantum chromodynamics transition predicted by the
  %standard model of particle physics,''
  Nature {\bf 443}, 675 (2006)
  [hep-lat/0611014].
  %%CITATION = HEP-LAT/0611014;%%

%\cite{pisarski-wilczek}
\bibitem{pisarski-wilczek} 
  R.~D.~Pisarski and F.~Wilczek,
  %``Remarks on the Chiral Phase Transition in Chromodynamics,''
  Phys.\ Rev.\ D {\bf 29}, 338 (1984).
  %%CITATION = PHRVA,D29,338;%%

%\cite{Cheng:2010fe}
\bibitem{Cheng:2010fe} M.~Cheng, S.~Datta, A.~Francis, J.~van der Heide,
  C.~Jung, O.~Kaczmarek, F.~Karsch and E.~Laermann {\it et al.},
  %``Meson screening masses from lattice QCD with two light and the strange
  %quark,''
  Eur.\ Phys.\ J.\ C {\bf 71}, 1564 (2011)
  [arXiv:1010.1216 [hep-lat]].
  %%CITATION = ARXIV:1010.1216;%%

\bibitem{multifractal}
A.~Rodriguez, L.J.~Vasquez, K.~Slevin and R.A.~R\"omer,
Phys.\ Rev.\ Lett.\ {\bf 105}, 046403 (2010); Phys.\ Rev.\ B {\bf 84},
134209 (2011). 

%\cite{Landa-Marban:2013oia}
\bibitem{Landa-Marban:2013oia} 
  D.~Landa-Marb\'an, W.~Bietenholz and I.~Hip,
  %``Features of a Simple IR Conformal Gauge Theory,''
  arXiv:1307.0231 [hep-lat];
  %%CITATION = ARXIV:1307.0231;%%
    
%\cite{Bietenholz:2013coa}
\bibitem{Bietenholz:2013coa} 
  W.~Bietenholz, I.~Hip and D.~Landa-Marb\'an,
  %``Microscopic Dirac Spectrum in a 2d Gauge Theory with Zero Chiral
  %Condensate,''
  arXiv:1310.3427 [hep-lat].
  %%CITATION = ARXIV:1310.3427;%%

%\cite{Golterman:2003qe}
\bibitem{Golterman:2003qe} 
  M.~Golterman and Y.~Shamir,
  %``Localization in lattice QCD,''
  Phys.\ Rev.\ D {\bf 68}, 074501 (2003)
  [hep-lat/0306002].
  %%CITATION = HEP-LAT/0306002;%%

%\cite{Greensite:2005yu}
\bibitem{Greensite:2005yu} 
  J.~Greensite, S.~Olejn\'ik, M.~Polikarpov, S.~Syritsyn and V.~Zakharov,
  %``Localized eigenmodes of covariant Laplacians in the Yang-Mills vacuum,''
  Phys.\ Rev.\ D {\bf 71}, 114507 (2005)
  [hep-lat/0504008].
  %%CITATION = HEP-LAT/0504008;%%

\end{thebibliography}
\end{document}